\documentclass[twocolumn]{aastex63}
\hypersetup{linkcolor=red,citecolor=blue,filecolor=cyan,urlcolor=black}

\newcommand{\kms}{km\,s$^{-1}$}
\newcommand{\OIII}{O\,{\scriptsize III}}

\newcommand{\CIII}{C\,{\scriptsize III}}

\newcommand{\NV}{N\,{\scriptsize V}}
\newcommand{\CIV}{C\,{\scriptsize IV}}
\newcommand{\HeII}{He\,{\scriptsize II}}

\newcommand{\hst}{\textit{HST}}

\newcommand{\spitzer}{\textit{Spitzer}}

\newcommand{\eazy}{\texttt{EAzY}}
\newcommand{\pypeit}{\texttt{PypeIt}}
\newcommand{\bagpipes}{\texttt{Bagpipes}}
\newcommand{\cgs}{erg\,s$^{-1}$\,cm$^{-2}$}
\newcommand{\cgsa}{erg\,s$^{-1}$\,cm$^{-2}$\,\AA$^{-1}$}

\usepackage{rotating}

\graphicspath{{./}{figures/}}

\shorttitle{MOSFIRE Spectroscopy of Fainter, $z\sim8$ IRAC-excess Galaxies}
\shortauthors{Roberts-Borsani et al.}

\begin{document}



\title{Nature and Nurture? Comparing Ly$\alpha$ Detections in UV-bright and Fainter [\OIII]+H$\beta$ Emitters at $z\sim8$ With Keck/MOSFIRE}

\correspondingauthor{Guido Roberts-Borsani}
\email{guidorb@astro.ucla.edu}

\author[0000-0002-4140-1367]{Guido Roberts-Borsani}
\affiliation{Department of Physics and Astronomy, University of California, Los Angeles, 430 Portola Plaza, Los Angeles, CA 90095, USA}

\author[0000-0002-8460-0390]{Tommaso Treu}
\affiliation{Department of Physics and Astronomy, University of California, Los Angeles, 430 Portola Plaza, Los Angeles, CA 90095, USA}

\author[0000-0002-3407-1785]{Charlotte Mason}
\affiliation{Cosmic Dawn Center (DAWN)}
\affiliation{Niels Bohr Institute, University of Copenhagen, Jagtvej 128, 2200 København N, Denmark}

\author[0000-0001-7782-7071]{Richard S. Ellis}
\affiliation{Department of Physics and Astronomy, University College London, Gower Street, London WC1E 6BT, UK}

\author[0000-0001-7459-6335]{Nicolas Laporte}
\affiliation{Kavli Institute for Cosmology, University of Cambridge, Madingley Road, Cambridge CB3 0HA, UK}
\affiliation{Cavendish Laboratory, University of Cambridge, 19 JJ Thomson Avenue, Cambridge CB3 0HE, UK}

\author[0000-0002-2772-8160]{Thomas Schmidt}
\affiliation{Department of Physics and Astronomy, University of California, Los Angeles, 430 Portola Plaza, Los Angeles, CA 90095, USA}

\author[0000-0001-5984-0395]{Marusa Bradac}
\affiliation{University of Ljubljana, Department of Mathematics and Physics, Jadranska ulica 19, SI-1000 Ljubljana, Slovenia}

\author[0000-0003-3820-2823]{Adriano Fontana}
\affiliation{INAF Osservatorio Astronomico di Roma, Via Frascati 33, 00078 Monteporzio Catone, Rome, Italy}

\author[0000-0002-8512-1404]{Takahiro Morishita}
\affiliation{Infrared Processing and Analysis Center, Caltech, 1200 E. California Blvd., Pasadena, CA 91125, USA}

\author[0000-0002-9334-8705]{Paola Santini}
\affiliation{INAF Osservatorio Astronomico di Roma, Via Frascati 33, 00078 Monteporzio Catone, Rome, Italy}

\begin{abstract}
The 100\% detection rate of Ly$\alpha$ emission in a sample of four luminous $z\sim8$ galaxies with red \spitzer/IRAC colors suggests objects with unusual ionizing capabilities that created early ionized bubbles in a neutral era. Whether such bubbles reflect enhanced ionizing properties (nature) or an overdense environment (nurture), however, remains unclear. Here we aim to distinguish between these hypotheses via a search for Ly$\alpha$ emission in five fainter galaxies drawn from the CANDELS-GOODS fields using a similar IRAC excess and UV magnitudes that should reflect reduced clustering effects. Using Keck/MOSFIRE we tentatively detect $>4\sigma$ line emission in only two targets at redshifts $z_{\rm Ly\alpha}$=7.1081 and $z_{\rm Ly\alpha}$=7.9622 with rest-frame EWs of 16-17 \AA, $\sim$1.5$\times$ weaker compared to their brighter counterparts. Thus we find a reduced rate for Ly$\alpha$ emission of $0.40^{+0.30}_{-0.25}$ compared to $1.00^{+0.00}_{-0.44}$ for more luminous examples. The lower rate agrees with predictions from simulations of a mostly neutral IGM and an intrinsic EW$_{0,\rm Ly\alpha}$ distribution for $z\sim6$ galaxies. However, even with an extreme EW$_{0,\rm Ly\alpha}$ model, it is challenging to match the detection rate for the luminous objects. SED-fitting of our fainter sample indicates young and star-forming systems, albeit with less extreme SFRs and ionization parameters compared to their luminous counterparts. The enhanced Ly$\alpha$ rate in luminous galaxies is thus likely a byproduct of both extreme ionizing properties as well as environment effects. Further studies with \textit{JWST} may be required to resolve the physical nature of this puzzling population.
\end{abstract}


\section{Introduction}
Pinpointing the birth of the first stars and galaxies and determining their role in transforming the intergalactic medium (IGM) from a neutral state to completely ionized remains one of the holy grails of observational cosmology and one of the main motivations for next generation telescopes (e.g., the \textit{James Webb} Space Telescope, Square Kilometre Array, and optical/IR Extremely Large Telescopes). A popular view is that reionization was driven largely by the abundant population of faint galaxies dominating the faint end of the galaxy luminosity function \citep{robertson15,robertson21}. However, the rapid change in the ionization state of the IGM at redshifts $6\lesssim z\lesssim8$ \citep{mason19,qin2021}, may suggest that rarer, more luminous sources provided a dominant contribution in concluding the process \citep[e.g.,][]{naidu20}. While the high neutral fraction of the IGM beyond $z\simeq$8 is well established \citep{treu13,mason19,hoag19,Bolan2022}, constraints on the sources that governed the reionization process are still sorely lacking: are their properties similar to those of galaxies found at $z\lesssim$7.5 or do they differ, perhaps reflecting an alternate evolutionary pathway as one approaches the first galaxies \citep[e.g.,][]{oesch14,mcleod16,oesch18}?

An important clue comes from spectroscopic observations of luminous galaxies with red \textit{Spitzer} Space Telescope IRAC 3.6-4.5 $\mu$m colors - a so-called ``IRAC excess'', generally attributed to intense
[\OIII]+H$\beta$ line emission polluting the IRAC 4.5 $\mu$m band. With rest-frame equivalent widths of EW$_{\rm 0}\gtrsim$1500 \AA, such line emission is suggestive of extreme sources with intense radiation fields, resulting in unusually high fractions of Ly$\alpha$ detections at $z\gtrsim$7.5 \citep{oesch15,zitrin15,rb16,laporte17,laporte17b,stark17,hashimoto18,endsley21,laporte21} in contrast with the much lower detection rates in magnitude-limited samples \citep[c.f.][]{fontana10,pentericci11,schenker14,mason18,mason19,hoag19}. Perhaps the most marked example was the recent spectroscopic follow up of a sample of four luminous ($H_{\rm 160}\sim25$ AB; $M_{\rm UV}\sim-22$ mag), IRAC-excess galaxies selected by \citet[][henceforth RB16]{rb16} in the CANDELS fields, which revealed Ly$\alpha$ emission in each of the four objects with Keck/MOSFIRE and VLT/X-Shooter spectroscopy \citep[RB16;][]{stark17,laporte17b}. The surprising 100\% detection rate of the line is strong evidence for early ionized bubbles in a predominantly neutral medium.

Given the especially luminous nature of the sources, whether these bubbles are the consequence of extreme ionizing properties (i.e., nature) and/or due to clustered environments typical of more massive galaxies (i.e., nurture) remains an open question. As examples of the former, \citet{endsley21} showed a clear correlation between the EW distribution of detected Ly$\alpha$ and the strength of [\OIII]+H$\beta$ (as indicated by their \textit{Spitzer}/IRAC excesses) in a sample of 22 UV-bright galaxies at $z\simeq7$, suggestive of intrinsic properties playing a key role. Furthermore, additional spectroscopy of a number of IRAC-excess sources revealed line emission from high-excitation UV lines (e.g., \NV$\lambda$1240, [\CIII],\CIII]$\lambda\lambda$1907,1909, and \HeII$\lambda$1640) suggestive of hard spectra, extreme radiation fields \citep{stark17} and possible non-thermal contributions \citep{laporte17b,mainali18}. The detection of these lines suggests this particular population of galaxies may harbor unusual properties and stellar populations that are able to more efficiently carve out large, early ionized bubbles and have enhanced Ly$\alpha$ production.

However, environmental effects likely also play a role. With three of the four luminous RB16 galaxies residing in the same CANDELS-EGS field, cosmic variance and a clustered environment may play a significant role: indeed, spectroscopic confirmations of fainter $z\simeq7-9$ satellite galaxies around two of the RB16 EGS objects indicates the likely presence of overdensities, with physical separations between the galaxies extending merely 0.7-3.5 physical Mpc (pMpc) along the line of sight \citep{tilvi20,larson22}. Furthermore, the recent photometric analysis by \citet{leonova21} using deep \hst\ $Y-$band and \spitzer/IRAC imaging revealed an enhancement of neighbouring galaxies around each of the three UV-bright EGS galaxies of the RB16 sample, by a factor of $\sim3-9\times$ compared to blank fields. Such analyses suggest a ubiquity of overdense regions around luminous, Ly$\alpha$-emitting IRAC-excess sources, although deep spectroscopic measurements are required for confirmation.

Thus, recent analyses have revealed evidence for both exceptional properties and clustered environments for these intriguing samples, making it challenging to discern the primary cause of the early ionized bubbles. One way to disentangle the effects of nature (intrinsic properties) and nurture (environment) is to extend such analyses to fainter galaxies that are less subject to the same clustering effects expected in predominantly luminous samples \citep{bn14,qiu18,harikane22,qin22}. As such, in this paper we seek to determine whether the fainter-end of the IRAC-excess galaxy population reside in ionized bubbles to the same extent as their luminous counterparts, via the presence of Ly$\alpha$. We therefore present the spectroscopic search for Ly$\alpha$ in five fainter ($M_{\rm UV}\sim-21$ mag) IRAC-excess sources with Keck/MOSFIRE, selected as part of the extended RB16 sample with identical data sets and thus minimizing selection effects and ideally suited for a one-to-one comparison with their luminous ($M_{\rm UV}\sim-22$ mag) counterparts. 

The paper is structured as follows. In Section~\ref{sec:photo} we describe the target selection and photometric data sets adopted, Section~\ref{sec:mosfire} describes our Keck/MOSFIRE observations and presents the results of those observations. Section~\ref{sec:lya_sims} compares the detection rates of Ly$\alpha$ across bright and fainter samples with Ly$\alpha-$IGM modelling, while Section~\ref{sec:sedfitting} derives and contrasts their intrinsic properties via SED-fitting. Finally, we present a summary and our conclusions in Section~\ref{sec:concl}. Where relevant, we assume \textit{H}$_{0}=$70 km/s/Mpc, $\Omega_{m}=$0.3, and $\Omega_{\wedge}=$0.7. All magnitudes are in the AB system \citep{oke83}.

\section{Target Selection \& Photometry}
\label{sec:photo}
\begin{figure}
\center
 \includegraphics[width=\columnwidth]{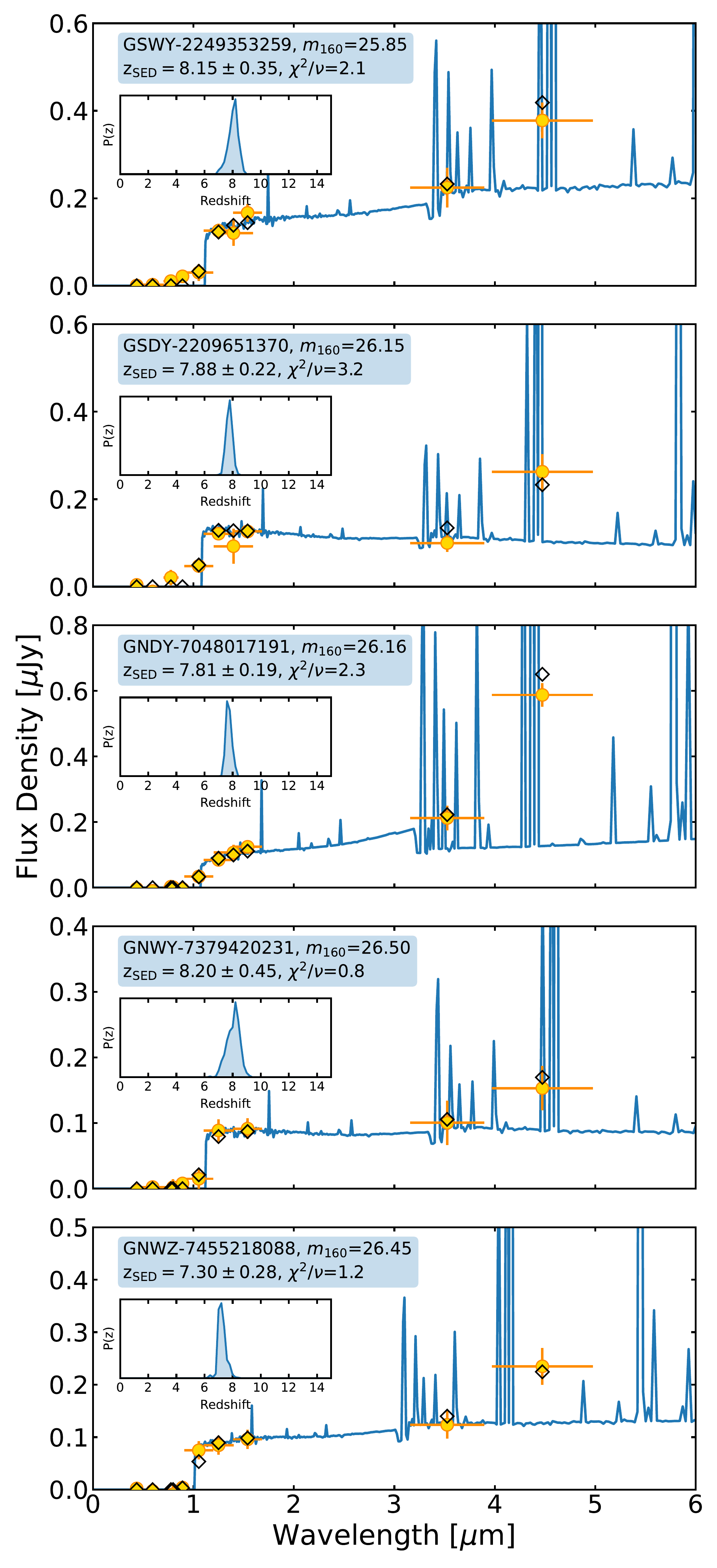}
 \caption{The best fit SEDs (blue lines) of our 5 primary targets, as well as their associated and modeled \hst\ and \spitzer/IRAC photometry (orange points with $1\sigma$ uncertainties and black diamonds, respectively). The inset plot shows the photometric redshift distribution resulting from the fit.}
 \label{fig:seds}
\end{figure}

To compare Ly$\alpha$ visibility statistics with the primary, luminous IRAC-excess sample of RB16 galaxies, we select a subset of the fainter galaxies presented in Table~5 of the same paper, which were selected in identical fashion to the bright sample - i.e., through an initial NIR color cut to constrain the location of the Lyman break, followed by photo-$z$ modelling with \eazy\ (see RB16 for details) - albeit with fainter magnitude limits. A total of 5 galaxies were selected here (GSWY-2249353259, GSDY-2209651370, GNDY-7048017191, GNWY-7379420231, and GNWZ-7455218088) based on their approximate $M_{\rm UV}$ values and visibility with the Keck telescopes. Three of the targets reside in the GOODS-North field, two in the GOODS-South field and all have $H_{\rm 160}\geqslant26.1$, [3.6]-[4.5]$\geqslant$0.5 and reported photometric redshifts of $z_{\rm phot}>7$ from \eazy. Each of the targets benefit from deep \hst\ and \spitzer/IRAC photometry, which we adopt from \citet{bouwens15}: the filters included as part of the data set are optical filters from \hst/ACS (F435W, F606W, F775W, F814W, F850LP), NIR filters from \hst/WFC3 (F098M, F105W, F125W, F140W, F160W) and IR filters from \spitzer/IRAC (CH1 and CH2) - for details on the construction of the data set and RB16 selection, we refer the reader to \citet{bouwens15} and RB16. We fit all of the available photometry for each object using the Bayesian Analysis of Galaxies for Physical Inference and Parameter EStimation \citep[\bagpipes;][]{carnall18} SED-fitting code, adopting an exponentially delayed star formation history. The choice of model is primarily motivated by previous studies highlighting the need for extended star formation to trace the bulk of the stellar mass formed as well as more recent and intense star formation to account for the boosting of [\OIII]+H$\beta$ nebular line emission in the \spitzer/IRAC bands (e.g., \citealt{stark17,rb20,laporte21}). The model and its parameter ranges are similar to those adopted by \citet{stark17} for characterization of the luminous RB16 IRAC-excess objects and thus allow for a close comparison of results. The free parameters were a stellar mass formed (log $M_{*}/M_{\odot}$) of [6,10], a star formation timescale ($\tau$) of [0,100] Gyrs, a metallicity ($Z/Z_{\odot}$) of [0,0.5], stellar ages of [0,0.1] Gyrs, dust attenuation ($A_{\rm v}$/mag) of [0,1] mag, and an ionization parameter (log\,$U$) of [$-4,-1$] with a velocity dispersion for all absorption and emission lines of 150 \kms. Redshifts were allowed to vary between $z=$[0,15]. The resulting photo-$z$ distributions and SEDs are plotted in Figure \ref{fig:seds}, where we find the photometric redshift solutions are virtually identical (and within errors) to those reported in Table~5 of RB16. We note that we perform a similar fit to the data with the addition of a bursty component with age 0-10 Myrs, mass formed (in log units) 6-10 $M_{*}/M_{\odot}$ and shared metallicity and dust components, however we find that the model is poorly constrained by the data and results in poorer fits. We summarize the list of targets, along with their basic photometric quantities, in Table~\ref{tab:props}. For comparison, we also compile the photometric properties of the luminous RB16 sample in the bottom half of Table~\ref{tab:props}. Our faint sample has a mean $H_{\rm 160}$ magnitude of 26.22$\pm$0.15 AB and absolute magnitude (assuming photometric redshifts and a UV slope of $\beta=-2$) of $-20.93\pm0.15$ mag, compared to 25.12$\pm$0.06 and $-22.02\pm0.06$ (assuming spectroscopic redshifts) for the bright sample. Adopting the $M_{\rm UV}-M_{h}$ relation (where $M_{h}$ denotes the dark matter halo mass) from \citet{mason15}, we estimate the $\Delta M_{\rm UV}\sim1$ mag difference to translate to the bright galaxies residing in halos $\sim3\times$ more massive than the fainter ones, and consistent with simulations showing $m_{\rm UV}<25.5$ AB galaxies residing in larger ionized bubbles \citep{qin22}.

\section{Keck/MOSFIRE Observations}
\label{sec:mosfire}
\subsection{$Y-$band Spectroscopy}
\label{subsec:observations}
For our spectroscopic campaign, we used the Multi-Object Spectrometer for Infra-Red Exploration \citep[MOSFIRE;][]{mclean12} on the Keck I telescope, targeting each of our sources with $Y-$band spectroscopy in search for Ly$\alpha$ emission. MOSFIRE slitmasks were constructed using the MOSFIRE Automatic GUI-based Mask Application (MAGMA) software and slit widths of 0.7$''$ (affording a spectral resolution of $R\sim3500$). Thanks to MOSFIRE's large FOV ($\sim60'\times30'$) the primary targets were multiplexed over a total of 2 slitmasks to increase observing efficiency - the primary targets included on Mask 1 were GSWY-2249353259 and GSDY-2209651370, while those included on Mask 2 were GNDY-7048017191, GNWY-7379420231 and GNWZ-7455218088. In both slitmasks, a bright nearby star was included in one of the slits for seeing- and throughput-monitoring purposes.

The observations were distributed over an approximate 7-month window and carried out over a total of 6 nights (5 full nights and 2 half nights; 24th November 2020, 25th November 2020, 7th December 2020, 20th April 2021, 21st April 2021, 16th May 2021 and 17th May 2021 - the latter two of which were half nights) using a $1\farcs25$ ABBA dither pattern - the precise amplitude of the dither was carefully checked to ensure the primary sources did not fall on any slit-contaminating sources. While most of the first half of 24th November 2020 was lost due to poor weather conditions and the entire night of 20th April 2020 lost due to cloud coverage, the rest of the observations were carried out in good weather conditions and sub-arcsec seeing ($\sim0.84''$, $\sim'0.87'$, $\sim0.73''$, $\sim0.63''$, $\sim0.67''$ and $\sim0.91''$ for 24th November 2020, 25th November 2020, 7th December 2020, 21st April 2021, 16th May 2021 and 17th May 2021, respectively). A summary of the targets and the telescope observations is provided in Table~\ref{tab:props}. In the case of Mask 1, it became apparent after the second night that the slit containing GSDY-2209651370 was too long ($\sim3'$), causing some curvature of the resulting sky lines at the opposite end of the target's position. As such, the slit was shortened to $\sim1.9'$ to avoid the issue further.

\begin{figure*}
\center
 \includegraphics[width=\textwidth]{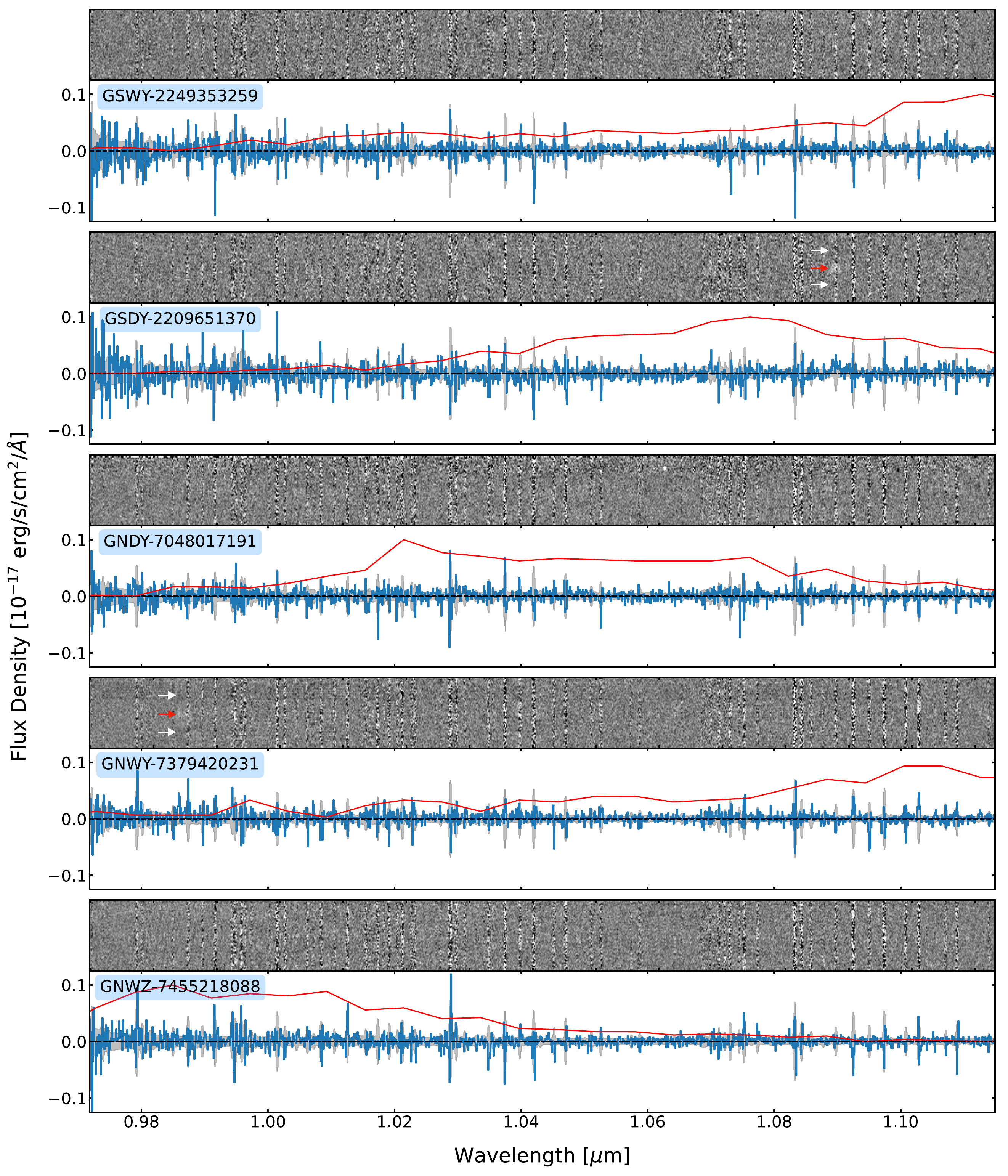}
 \caption{The 2D (top) and 1D (bottom) Keck/MOSFIRE spectra of each of our 5 targets in the GOODS-S and GOODS-N fields. Each of the spectra were reduced with the \pypeit\ data reduction software. In each case, red and white arrows in the top panels mark the positive and negative traces of probable Ly$\alpha$ detections. Blue lines, grey-shaded regions, and red lines in the bottom panels denote the extracted 1D spectra, associated $1\sigma$ uncertainties, and the \bagpipes-derived Ly$\alpha$ redshift probability function, respectively.}
 \label{fig:spectra}
\end{figure*}

The data were reduced using the \pypeit\ data reduction package \citep{pypeit1,pypeit2}, which uses telescope/instrument-specific data reduction scripts invoked from an input configuration file (henceforth referred to as the ``\pypeit\ file''). The \pypeit\ file takes as input the science and calibration files to reduce, as well the parameters required to reduce, calibrate and co-add the data. We adopt the default Keck/MOSFIRE parameters
\footnote{\url{https://pypeit.readthedocs.io/en/release/pypeit_par.html}} and reduce the data in ABBA sequence blocks (i.e., for a given sequence, A-B and B-A subtraction is performed for background subtraction before co-adding AB frames and BA frames, resulting in two background-subtracted images), using the spectral trace of a bright star in one of the slits on each mask as a position reference. The co-added science exposures were used for wavelength calibration and tilt measurements, while flat frames were used for slit edge tracing in addition to flat-fielding. We note that in the case of Mask 1 (i.e., the GOODS-S targets), we used only the flats of 7th December 2020, since this allowed us to artificially define the slit traces of that night onto the data taken on 24th and 25th November 2020, thereby providing us with a significantly improved wavelength solution based on the smaller slit adopted for the December observations. The reduced A-B (or B-A) data frames were subsequently co-added using the extracted 1D spectra from each frame as weights, and flux-calibrated using a sensitivity function (i.e., the ratio of $F_{\lambda}$ flux density to electron count density) generated from an archived, flux-calibrated standard star spectrum from ESO, providing us with fully reduced, calibrated and co-added 2D and 1D spectra which we display in Figure \ref{fig:spectra}. The total exposure times resulting from the observations and data reduction were $\sim$6.3 hours for the GOODS-S mask and $\sim$11.8 hours for the GOODS-N mask.

\subsection{Detections of Ly$\alpha$ at $z>7$}
\begin{figure*}
\center
 \includegraphics[width=\textwidth]{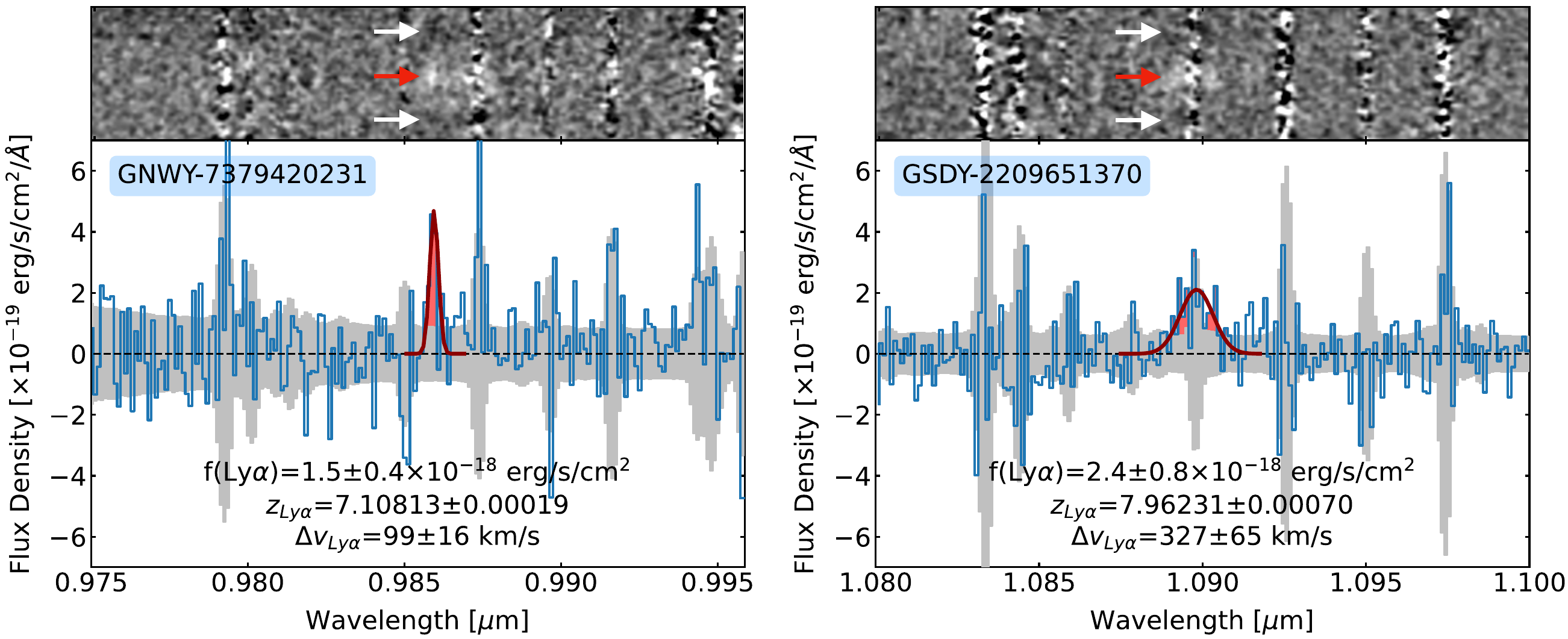}
 \caption{Tentative Ly$\alpha$ detections for two of our targeted galaxies: GNWY-7379420231 (left) and GSDY-2209651370 (right). Each 2D spectrum is smoothed by a Gaussian kernel and} shown in the top panel, with red and white arrows to mark the positive and negative Ly$\alpha$ traces respectively, while the bottom panels show the collapsed 1D spectra (blue line), the associated 1$\sigma$ uncertainty array (grey fill), and the best fit Gaussian model (dark red line) to each line (red fill).
 \label{fig:lya}
\end{figure*}

Each of the resulting 1D and 2D $Y-$band spectra presented in Figure \ref{fig:spectra} were visually (and independently) inspected by two authors (GRB and NL) to search for Ly$\alpha$ emission. Tentative emission lines were found in two of the targeted galaxies, namely GNWY-7379420231 and GSDY-2209651370, and are shown in Figure \ref{fig:lya}. 

For GNWY-7379420231, a single, narrow emission line is clearly visible in the 2D spectrum, showing a positive, central component with two negative counterparts above and below it separated by approximately $\pm$14 pixels, which corresponds to our dither choice of 1.25$''$. The line resides between sky lines and a simple Gaussian fit (ignoring potential line asymmetries due to a neutral IGM) finds the central wavelength to lie at 9859.5 \AA\ with an integrated signal-to-noise ratio (SNR) of $\sim4.5\sigma$ (using the integral of the fitted Gaussian and noise measured in sky line-free regions of the spectrum either side of the line). No other convincing emission line is seen in the rest of the 1D or 2D spectra. To ensure the line is not due to a spurious bad frame or artefact in the data, we first visually inspect and measure the S/N ratio at the position of the line in all individual 1D spectra (integrating over the spectrum at the position of the original Gaussian and again adopting noise from the same sky line-free regions used in the full stack) from the AB and BA exposure pairs, finding no single exposure where a significant peak is found. To add statistical robustness and asses whether the line is consistently found in randomly-sampled portions of the data, we adopt a Monte Carlo approach where we randomly sample (without replacement) 50\% of all AB and BA exposure pairs and 2D co-add those selected spectra before re-extracting the 1D spectrum and re-estimating the SNR of the apparent line. Repeating this 100 times, we find that 67\% of the iterations reveal a $>3.2\sigma$ (scaled from 4.5$\sigma$) line, while this increases to 73\% (at $>3.9\sigma$) if we perform the same simulations with 75\% of the data. The detection rates lie above those expected for a spurious line in 1-2 bad frames ($\sim$39\% and $\sim$64\% when using half or three quarters of the full data set, respectively), but not significantly enough to make this test conclusive.

A similar case is found for GSDY-2209651370, which displays a single, broad emission line: the line itself falls behind a sky line, however it is sufficiently broad to be clearly visible either side of the sky line and a positive/negative trace pattern similar to the line in GNWY-7379420231 is seen at the expected positions. Again, no other line is found in the rest of the 1D or 2D spectra. Masking the sky line and fitting another simple Gaussian profile, we find the emission line has a central wavelength of 10898.0 \AA\ and an integrated SNR of $\sim5.8\sigma$. Given its location behind a sky line, we cannot reliably extend the above statistical analyses to GSDY-2209651370, since measurements at the center of the line would be contaminated by the overlapping sky line, and the ``wings'' of the line either side of the sky line would be too faint for any reliable measurements.

Both of the identified lines are consistent with Ly$\alpha$ emission at $z>7$ and the hypothesis of strong [\OIII]+H$\beta$ emission lines as the cause of the especially red \spitzer/IRAC band (see \citealt{rb20} for a description of Balmer break contributions to the red colors at $z\gtrsim7.5$, however). As such, we tentatively confirm GNWY-7379420231 and GSDY-2209651370 at redshifts of $z_{\rm Ly\alpha}=7.108$ and $z_{\rm Ly\alpha}=7.962$, respectively, consistent with their $P(z)$'s estimated with \bagpipes. We caution, however, that considering the detection rates estimated above for GNWY-7379420231 and the faintness of the lines, we cannot conclusively rule out a spurious nature for either, and thus ensuing detection rates should effectively be regarded as upper limits.

Correcting for instrumental broadening with a Gaussian fit to a nearby sky line, we find the emission line of GNWY-7379420231 to be just unresolved ($\delta\sigma_{\rm gauss}=0.24$ \AA) and thus assume a maximum FWHM given by the sky line. This is not the case for GSDY-2209651370, whose emission line is significantly wider. The instrumentally-corrected Gaussian fits to the lines yield peak fluxes of 4.7$\pm$1.6$\times10^{-19}$ \cgsa\ and 2.0$\pm$0.5$\times10^{-19}$ \cgsa, integrated fluxes of 1.5$\pm$0.7$\times10^{-18}$ \cgs\ and 2.4$\pm$0.8$\times10^{-18}$ \cgs, and line FWHMs of 3$\pm$1 \AA\ and 12$\pm$3 \AA\ (corresponding to $\Delta v_{\rm Ly\alpha}=99\pm32$ \kms\ and $\Delta v_{\rm Ly\alpha}=340\pm79$ \kms) for GNWY-7379420231 and GSDY-2209651370, respectively. Assuming a flat, underlying continuum given by the $H_{\rm 160}$ photometry, these would correspond to rest-frame equivalent widths of EW$_{\rm 0,Ly\alpha}=16\pm8$ \AA\ and EW$_{\rm 0,Ly\alpha}=17\pm6$ \AA.

For non-detections of Ly$\alpha$ in GSWY-224935325, GNDY-7048017191, and GNWZ-7455218088, we place upper limits on the integrated flux and EW of the line. For the former, we take the median noise value in an uncontaminated (i.e., free of sky lines) portion of the noise spectrum, close to the photometric redshift of each galaxy and multiply the resulting value by the square-root of the number of pixels considered. For the latter, we use the resulting upper limit on the integrated flux and divide by the $H_{\rm 160}$ continuum as well as a (1+$z_{\rm phot}$) factor as above to obtain the upper limit on the rest-frame EW. The derived range of values is $<$6-9 \AA, and we summarize these properties in Table \ref{tab:props}. We note here that we do not account for slit loss effects.

\section{Comparing Lyman-$\alpha$ Detection Rates with Models}
\label{sec:lya_sims}
Here we aim to place the detection rates found for the UV-bright ($<M_{\rm UV}>\,=-22.02$ mag) and fainter ($<M_{\rm UV}>\,=-20.93$ mag) IRAC-excess samples into context: how do the rates compare, and how well (or poorly) are they matched to simulations of a neutral intergalactic medium and an assumed EW$_{\rm 0, Ly\alpha}$ distribution function? Quantifying the first of the two points determines whether any apparent differences are statistically significant, while the second determines whether our understanding of the $z\sim8$ IGM and underlying physics of typical galaxy populations can explain the derived detection rates - the models account for clustering effects (i.e., the nurture effect), and thus allow us to quantify whether a change in the intrinsic properties of galaxies (i.e., the nature effect) is also needed.

For the first point, out of our five faint $z>7$ IRAC-excess sources targeted with Keck/MOSFIRE, we find tentative Ly$\alpha$ detections in only two sources. Using the small number statistical limits presented by \citet{gehrels86}, this places the Ly$\alpha$ detection rate in our sample at 0.40$_{-0.25}^{+0.30}$ (assuming 16\% and 84\% binomial percentiles). While the small number statistics prevent us from claiming statistically significant differences, the detection rate found here suggests differences with the 1.00$^{+0.00}_{-0.44}$ detection rate of Ly$\alpha$ in the primary, luminous sample of RB16 determined by \citet{stark17}. The low intensities, integrated line fluxes and small EWs of the probable lines also make these among the faintest and weakest Ly$\alpha$ emission found thus far at comparable redshifts (c.f. \citealt{finkelstein15,oesch15,zitrin15,song16,laporte17,hashimoto18}): both of our lines display rest-frame EWs of 16-17 \AA, well below the Ly$\alpha$ emitter threshold of EW$_{\rm 0}>25$ \AA\ generally assumed for inclusion as LBGs in reionization-constraining studies, and a median factor of $\sim1.5\times$ weaker than the EWs reported for their luminous IRAC-excess counterparts.

On the second point, the flux of Ly$\alpha$ we observe is the product of the intrinsic Ly$\alpha$ flux that emerges from the ISM and the fraction that is transmitted through the reionizing IGM. The transmitted fraction is thus a function of the integrated optical depth of Ly$\alpha$ through the IGM, determined by the distribution of neutral gas due to reionization, and the line shape of Ly$\alpha$ determined from radiative transfer of the line through the galaxy ISM. To interpret our observations we compare our Ly$\alpha$ detection rate to that expected from inhomogeneous reionization simulations using a Monte Carlo simulation. For each IRAC-selected galaxy we generate 1000 mock Ly$\alpha$ observations by simulating a line EW from an underlying distribution and converting the EW to a corresponding line flux. We then compare to the reported flux uncertainties (from Table~\ref{tab:props}) to classify the mock observations as a detection or not. The Ly$\alpha$ EWs are sampled from the model EW distributions by \citep{Mason2018a,mason18} which are a function of UV magnitude and the IGM neutral fraction $p(W | M_\textsc{uv}, \overline{x}_\textsc{hi})$, constructed by forward-modelling an assumed intrinsic Lya EW distribution through the realistic inhomogeneous reionization Evolution of Structure simulations over thousands of sightlines \citep{Mesinger2016}. In this way we account for the expected UV magnitude dependence of Ly$\alpha$ transmission during reionization, whereby the brightest galaxies live in overdensities that should reionize early and thus have higher Ly$\alpha$ transmission \citep[e.g.,][]{Mason2018a,mason18,qin2021,leonova21}.

For each galaxy, we sample a UV magnitude, and redshift, from within the $1\sigma$ uncertainties in Table~\ref{tab:sed_props} (see Section \ref{sec:sedfitting}) and Table~\ref{tab:props}, respectively assuming Gaussian errors. We then sample the IGM neutral fraction at the spectroscopic or sampled photometric redshift from the constraints on the IGM timeline by \citet[][]{Mason2019b}, which conservatively includes only the \citet{planck20} electron scattering optical depth and quasar dark pixel fraction \citep{McGreer2015}. For example, at $z\sim8$ the inferred IGM neutral fraction is $\overline{x}_\textsc{hi} = 0.73_{-0.27}^{+0.15}$. We then sample from the corresponding Ly$\alpha$ EW distribution $p(W | M_\textsc{uv}, \overline{x}_\textsc{hi})$ by \citet{Mason2018a}. For each realization we draw a Ly$\alpha$ EW and forward-model the Ly$\alpha$ emission line using the continuum flux density derived using $M_{\rm UV}$ and $\beta$ slopes from Table~\ref{tab:sed_props}, we then sample a Ly$\alpha$ velocity offset from the model by \citet{Mason2018a} and set FWHM equal to the velocity offset following \citet{Verhamme2015}. We then perform a mock observation summing the flux over a wavelength region with width $2\times$FWHM, centered on the peak ``observed'' Ly$\alpha$ wavelength, and summing the noise from our MOSFIRE observations in quadrature in that wavelength range.
Simulated fluxes above the $2\sigma$ limit (we choose a $2\sigma$ threshold to mimic the lowest significance of the detected line EWs in Table~\ref{tab:props}) given the galaxy's observed flux uncertainty is classified as a detection. We repeat this process 1000 times for each galaxy to build a probability distribution for the expected number of detections.

\begin{figure*}
\center
 \includegraphics[width=0.85\textwidth]{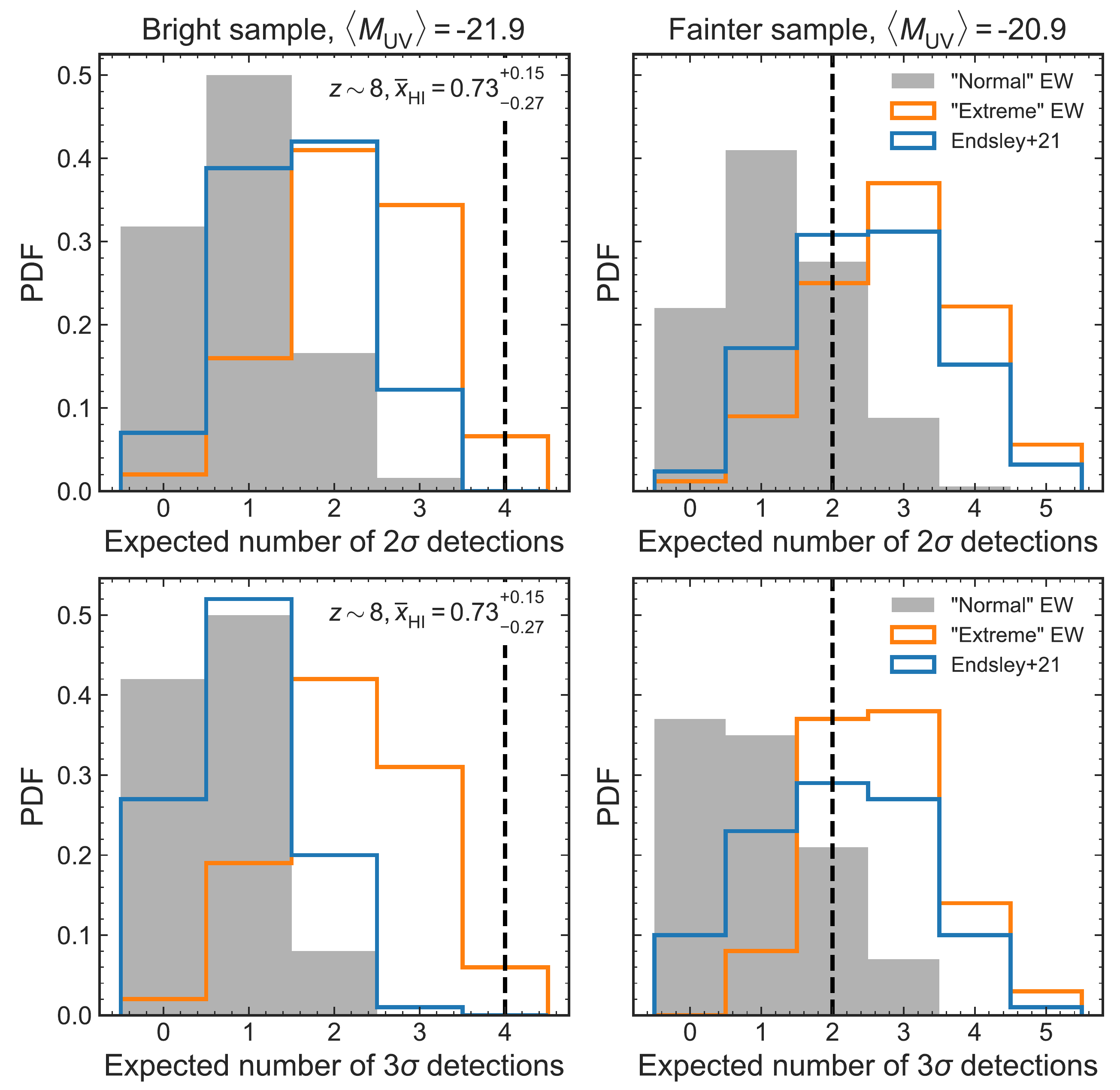}
 \caption{
 Probability distribution of expected number of $2\sigma$ (top panels) and $3\sigma$ (bottom panels) Ly$\alpha$ detections for galaxies with the same observed photometric properties, redshifts and flux uncertainties as in Table~\ref{tab:props} (left panels for the UV-bright sample, right panels for the UV-fainter sample), forward-modelled using inhomogeneous reionizing IGM simulations \citep{mason18}. For each galaxy we sample the IGM neutral fraction at their spectroscopic redshift (or marginalize over photometric redshift) from the model-independent constraints on the reionization timeline by \citet{Mason2019b}, which includes only the CMB electron scattering optical depth \citep{planck20} and the dark pixel fraction in the spectra of $z\gtrsim5.5$ quasars \citep{McGreer2015}. We use three intrinsic Ly$\alpha$ EW distributions: ``normal'' (grey shaded region) -- the standard model as a function of UV magnitude and IGM neutral fraction by \citet{mason18}, ``extreme'' (orange line) -- identical to the ``high EW'' model by \citep{mason18}, where these galaxies are given the same intrinsic EW distribution as $M_\textsc{uv} \sim -16$ galaxies, which typically have stronger emitted EWs; and the distribution observed by \citet{endsley21} for $1-6L_\star$ galaxies at $z\sim6$ (blue line). Black dashed lines show the observed number of $\geq2\sigma$ detections. The number of detections in our fainter sample is consistent with all of our models for Ly$\alpha$, however, the bright sample is challenging to explain without requiring these galaxies have very extreme intrinsic Ly$\alpha$ emission. We note that should the tentative Ly$\alpha$ lines turn out to be spurious, the 0\% detection rate in the fainter population would merely reinforce the contrast between the two galaxy samples and strengthen our conclusions.
 }
 \label{fig:lya_igm}
\end{figure*}

In Figure~\ref{fig:lya_igm} we show the results of these simulations. The left panels shows the probability distribution of the expected number of (EW) detections for the UV-bright IRAC-selected sample by RB16, and the right panels shows the same for the fainter sample presented here - we report the results for the expected number of $2\sigma$ and $3\sigma$ detections in the top and bottom panels of the figure, respectively. The grey shaded region shows the expected number of detections for the ``normal'' $p(W | M_\textsc{uv}, \overline{x}_\textsc{hi})$ EW distribution model by \citet{Mason2018a}, which assumes an intrinsic Ly$\alpha$ EW distribution from an empirical model built using $-21 \lesssim M_\textsc{uv} \lesssim -16$ Lyman break galaxies at $z\sim6$, which were not selected by strong IRAC excesses \citep{Debarros17,Fuller2020,Bolan2022}.

As discussed by \citet{mason18} this intrinsic distribution fails to match the observed frequency of detections in the bright IRAC-excess galaxies, implying that these galaxies have enhanced Ly$\alpha$ emission compared to non IRAC-excess galaxies, or enhanced IGM transmission that is not accounted for in current simulations (as these galaxies are situated already in the most transmissive, early reionized regions in our simulations), or a combination of these factors. We thus also predict the expected number of detections for two additional intrinsic Ly$\alpha$ EW models. The first is an ``extreme'' model where these galaxies are given the intrinsic EW distribution for UV-faint galaxies, which typically have higher EW at lower redshifts \citep[this is identical to the ``high EW'' model by][]{mason18}. The second is an EW distribution for $>L_\star$ galaxies at $z\sim6$, including a number with strong IRAC-excesses, by \citet{endsley21} which may be more appropriate for our sample. We find the bright galaxies challenging to explain with any of our models. Only our ``extreme'' EW model has some probability of all 4 galaxies being detected. We ran an additional simulation in a fully ionized universe and found that, even then, detecting Ly$\alpha$ in 4/4 UV-bright galaxies was unlikely given our intrinsic EW models. On the other hand, we find that our fainter sample is consistent within $1\sigma$ with all three intrinsic EW models, and most consistent with the ``high EW'' and \citet{endsley21} distributions. Such a result suggests that these fainter IRAC-excess galaxies are fully consistent with current constraints on reionization, though their implied ionizing properties may favour enhanced Ly$\alpha$ emission compared to galaxies not selected with an IRAC excess (the ``normal'' model). 

From our derived detection rate, measured line strengths, and Ly$\alpha$ simulations (which account for both intrinsic galaxy physics and clustering effects), we conclude that the brightest IRAC-excess galaxies likely transmit enhanced levels of Ly$\alpha$ due to both (i) very high intrinsic EWs ($\gtrsim 100$\,\AA) and (ii) residing in the most overdense, early reionized regions. Their fainter IRAC-selected counterparts, on the other hand, do not emit such strong Ly$\alpha$, likely due to having less intense ionizing radiation fields, and are consistent with models where they reside in less overdense environments ionized by $<L_{*}$ galaxies. We note that in a case of zero Ly$\alpha$ detections in the fainter sample, the contrast between the two samples would only increase and our conclusions reinforced. Should GNWY-7379420231 and GSDY-2209651370 be revealed as low-$z$ interlopers, our conclusions would remain unchanged but at lower statistical significance.

\section{Are Bright and Fainter IRAC-Excess Galaxies Intrinsically Different? Clues from SED Modelling}
\label{sec:sedfitting}
Quantifying the gaseous and stellar properties of our sample of fainter galaxies and comparing to those of the UV-bright IRAC-excess galaxies allows us to assess whether the differences in their Ly$\alpha$ detection rates (quantified and discussed in Section~\ref{sec:lya_sims}) are attributable to differences in physical properties (i.e., nature). While the fainter objects do not possess the additional $J-$ and $H-$band MOSFIRE spectra used by e.g., \citet{stark17} and \citet{mainali18}, or longer-wavelength NIR X-Shooter data by \citet{laporte17b}, to constrain the radiation fields of the UV-bright RB16 galaxies through detections (or upper limits) of \NV, \CIV, \HeII, and [\CIII] line emission, it is nevertheless instructive to determine and contrast their properties using the deep photometry available and new spectroscopic redshifts. We thus update the best-fit models for GNWY-7379420231 and GSDY-2209651370 by incorporating their spectroscopic redshifts and re-running \bagpipes\ as described in Section \ref{sec:photo}. Additionally, we also run our SED-fitting procedure over the luminous galaxies using the photometry presented in RB16.

We find in all cases the photometry of our faint objects is well fit by the models and reveal underlying properties within expected ranges, with stellar mass values of log\,M$_{*}$/M$_{\odot}=[8.55,9.25]$, star formation rate (SFR) values of log\,SFR/M$_{\odot}\, \rm yr^{-1}=[0.60,1.32]$, specific SFR values of log\,sSFR/yr$^{-1}=[-8.00,-7.93]$, stellar ages of $[4,62]$ Myrs, metallicities of $Z/Z_{\odot}=[0.24,0.28]$, dust obscuration of A$_{\rm v}=[0.28,0.94]$, and ionization parameters of log\,$U=[-2.38,-1.63]$. We calculate absolute UV magnitudes and slopes directly from the best-fit SED - i.e., we adopt a bootstrap method where, for a sample of 1000 SEDs extracted using the posterior distributions of the \bagpipes\ free parameters, we fit the flux density between (rest-frame) wavelengths of 1300-2100 \AA\ with a simple power law to measure $\beta$. Simultaneously, we measure $M_{\rm UV}$ assuming the flux density at 1600 \AA\ and distance modulus given by the spectroscopic (or photometric) redshift. Such an approach results in UV slopes of $\beta=[-2.35,-1.65]$ and absolute UV magnitudes of $M_{\rm UV}=[-21.22,-20.27]$.

Our reported values - shown in Table~\ref{tab:sed_props} - are broadly consistent with those determined by \citet[][see their Table 3]{stark17} for EGS-zs8-1, EGS-zs8-2 and COS-zs7-1 (referred to as COSY here), in that they indicate more extreme systems than more representative samples of $z\sim8$ galaxies (e.g., \citealt{strait20,rb22}), with generally young stellar populations, low dust and metal contents, high ionization parameters, blue UV slopes, and elevated specific SFR values. Interestingly, we also observe some hints of a dichotomy between some properties of the bright and faint IRAC-excess samples. With ranges of log\,SFR/$M_{\odot}\,yr^{-1}$=[1.05,1.63], $\beta=[-2.2,-1.94]$, and log\,$U=[-1.79,-1.58]$, the bright sample displays enhanced SFRs, bluer UV slopes, and larger ionization parameters, indicative of more extreme star-forming systems that are likely to have boosted Ly$\alpha$ emission and enhanced ionizing capabilities to carve out early ionized bubbles. Although spectroscopic confirmations are required to validate such interpretations (through emission line diagnostics and continuum measurements), the distinction in photometric properties found here suggests the physics of the two populations may differ and provide some explanation for the differing detection rates of Ly$\alpha$.

Of course, the visibility of Ly$\alpha$ also depends on its velocity offset relative to the systemic redshift of the host galaxy, and thus whether it is sufficiently offset to be shifted out of resonance. As shown by several works (e.g., \citealt{erb14,stark15,Mason2018a}), the observed velocity offset is empirically correlated with galaxy UV luminosity at fixed redshift, implying UV-bright objects display larger Lya offsets compared to their fainter counterparts and thus enhanced Lya detection fractions. While we cannot quantify these parameters directly, we obtain predictions for the two IRAC-excess samples based on the $z\sim7$, $\Delta v_{\rm Ly\alpha}-M_{\rm UV}$ relation derived by \citet{Mason2018a}. Although the scatter in the literature data is large, we find a median velocity offset of $\sim165$ km/s for the fainter sample of galaxies, consistent with the lower end of values reported in the literature. Although uncertain, the median value is significantly lower than the 340-545 km/s values (a factor of $\sim2.0-3.3\times$) reported for the EGS objects in the luminous sample with Keck and \textit{JWST}/NIRSpec \citep{stark17,tang23}. The low value also contrasts with those reported for other UV-luminous galaxies displaying Ly$\alpha$ and additional lines through ground-based or \textit{JWST}/NIRSpec observations (e.g., \citealt{willott15,tang23}), by similar factors. The contrasting Ly$\alpha$ velocity offsets between the two IRAC-excess populations suggests these could be playing a significant role in the transmission of the line through the neutral IGM.

Direct verification of the Ly$\alpha$ offsets in the fainter galaxies is required to confirm such a contrast, and observations (spectroscopic and photometric) with \textit{JWST} (e.g., GO 1747, ERS 1324, ERS 1345, and GO 2279) will confirm these tantalizing clues by providing the rest-frame UV, optical, and IR diagnostics \citep{rb21} required to disentangle intrinsic galaxy physics (nature) from environmental (nurture) effects and determine the primary cause of the early ionized bubbles.

\section{Summary \& Conclusions}
\label{sec:concl}
We present deep Keck/MOSFIRE spectroscopy over five sources in the GOODS fields, selected on account of their \spitzer/IRAC excesses, intrinsically fainter ($M_{\rm UV}\sim-21$ mag) magnitudes to minimize clustering effects, and $z\sim8$ photometric redshifts. We tentatively confirm two of our sources with $>4\sigma$ Ly$\alpha$ emission at $z_{\rm Ly\alpha}=7.10813$ and $z_{\rm Ly\alpha}=7.9622$ and, through a comparison of Ly$\alpha$ detection rates, detailed simulations of IGM$-$Ly$\alpha$ opacity, and characterization of galaxy physics with SED-fitting, we offer a comparison in the potential of these faint galaxies and their more UV luminous counterparts to carve out early ionized bubbles in the Epoch of Reionization. Our findings can be summarized as follows:

\begin{itemize}
    \item We determine the detection rate of Ly$\alpha$ in our fainter galaxies to be 0.40$^{+0.30}_{-0.25}$, compared to 1.00$^{+0.00}_{-0.44}$ for the luminous sample of RB16 selected in identical fashion. The rest-frame EWs of our detections are 16-17 \AA, marking some of the weakest Ly$\alpha$ detections yet at these redshifts, and we place strong (3$\sigma$) upper limits of 6-9 \AA\ on the rest of our sources using sky line-free regions of the spectrum. In the event that the tentative lines turn out to be spurious, the resulting 0\% detection rate in the fainter sample would increase the contrast between the two samples and strengthen the following conclusions.

    \item We find simulations of Ly$\alpha$ detection rates in a mostly neutral medium are able to match the observed fainter sample detection rate with a rest-frame EW$_{0,Ly\alpha}$ distribution characterized by ``normal'' galaxies selected with or without an IRAC-excess, while we are unlikely to reproduce the detection rate of their luminous counterparts even with the most extreme EW distribution, suggesting boosted Ly$\alpha$ emission in the latter sample.
    
    \item A comparison of galaxy properties from photoionization models shows the UV-fainter sample is potentially governed by less extreme star formation and reduced ionization parameters compared to their bright counterparts.

\end{itemize}

In this study we confirm the special nature of luminous, IRAC-excess galaxies by investigating whether their fainter counterparts - who are less subject to clustering and environmental effects - also reside in early ionized bubbles to comparable extent, via the detection of Ly$\alpha$ in a predominantly neutral medium. The contrasting detection rates and photometric properties found here suggest the more extreme systems of UV-bright galaxies are able to produce and emit larger amounts of Ly$\alpha$ photons and more efficiently carve out large, early ionized bubbles. However, whether the primary cause of this is their enhanced ionizing properties or the collective ionizing output from neighboring galaxies remains an open question. The arrival of \textit{JWST} and its unprecedented spectroscopic and photometric capabilities will provide the required measurements of emission line diagnostics (e.g., rest-frame UV lines such as Ly$\alpha$, [\CIII], \HeII, and \CIV, as well as rest-frame optical lines such as [\OIII]\,$\lambda$5007 \AA\ and H$\beta$) as well as confirmations of clustered environments. In addition, the expected confirmation of both existing and new $z>7$ samples from the telescope will add statistical robustness to the interpretations presented here, and settle the extent to which luminous IRAC-excess objects drive an early reionization.

\acknowledgments
GRB and TT acknowledge financial support from NASA through grants JWST-ERS-1342 and HST-GO-13459. CAM acknowledges support by the VILLUM FONDEN under grant 37459 and the Danish National Research Foundation through grant DNRF140. RSE acknowledges financial support from European Research Council Advanced Grant FP7/669253. NL acknowledges support from the Kavli fundation. MB acknowledges support by the Slovenian national research agency ARRS through grant N1-0238.

GRB extends his thanks to Rychard Bouwens for providing the photometry of the samples, as well as Joe Hennawi, Debora Pelliccia, and Caitlin Casey for their help with the \pypeit\ software. Additional thanks are extended to Adam Carnall for useful discussions regarding the interpretation of the \bagpipes\ modelling.

\bibliography{sample63}{}
\bibliographystyle{aasjournal}

\begin{sidewaystable}[htb]
\centering
\begin{tabular}{lccccccccccc}
\hline
ID & RA & DEC & $H_{\rm 160}$ & [3.6]-[4.5] & $z_{\rm phot}$ & $z_{Ly\alpha}$ & f(Ly${\alpha}$) & FWHM & EW$_{0,\rm Ly\alpha}$ & Mask ID & iTime\\
 & [J2000] & [J2000] & [AB] & [AB] & & & [$\times10^{-18}$ \cgs] & [\AA] & [\AA] & & [s]\\
\hline
GSWY-2249353259 & 03:32:24.93 & -27:53:26.04 & 25.85$\pm$0.13 & 0.57$\pm$0.25 & 8.15$\pm$0.35 & --- & $<$0.18(3$\sigma$) & --- & $<$6(3$\sigma$) & Mask1 & 22546\\
GSDY-2209651370 & 03:32:20.96 & -27:51:37.06 & 26.14$\pm$0.12 & 1.05$\pm$0.28 & 7.88$\pm$0.22 & 7.962 & 2.39$\pm$0.83 & 12$\pm$3 & 17$\pm$6 & Mask1 & 22546\\
GNDY-7048017191 & 12:37:04.81 & +62:17:18.98 & 26.16$\pm$0.07 & 1.10$\pm$0.20 & 7.81$\pm$0.19 & --- & $<$0.15(3$\sigma$) & --- & $<$6(3$\sigma$) & Mask2 & 42587\\
GNWY-7379420231 & 12:37:37.94 & +62:20:22.82 & 26.50$\pm$0.19 & 0.46$\pm$0.44 & 8.20$\pm$0.45 & 7.108 & 1.50$\pm$0.71 & 3$\pm$1 & 16$\pm$8 & Mask2 & 42587\\
GNWZ-7455218088 & 12:37:45.52 & +62:18:08.87 & 26.44$\pm$0.21 & 0.70$\pm$0.28 & 7.30$\pm$0.28 & --- & $<$0.24(3$\sigma$) & --- & $<$9(3$\sigma$) & Mask2 & 42587\\
\hline
EGSY8p7$^{[1]}$ & 14:20:08.50 & +52:53:26.60 & 25.26$\pm$0.09 & 0.76$\pm$0.14 & --- & 8.683 & 17.0$\pm$7.5 & 11$\pm$8 & 28$\pm$13 & --- & --- \\
EGS-zs8-1$^{[2]}$ & 14:20:34.89 & +53:00:15.35 & 25.03$\pm$0.05 & 0.53$\pm$0.09 & --- & 7.730 & 17.0$\pm$3.0 & 13$\pm$3 & 21$\pm$4 & --- & --- \\
EGS-zs8-2$^{[3]}$ & 14:20:12.09 & +53:00:26.97 & 25.12$\pm$0.05 & 0.97$\pm$0.18 & --- & 7.477 & 7.4$\pm$1.0 & --- & 9$\pm$1 & --- & --- \\
COSY$^{[3]}$ & 10:00:23.76 & +02:20:37.00 & 25.06$\pm$0.06 & 1.03$\pm$0.15 & --- & 7.154 & 25.0$\pm$4.0 & --- & 28$\pm$4 & --- & --- \\
\hline
\end{tabular}
\caption{The spectro-photometric properties of our sample of intrinsically fainter Keck/MOSFIRE targets (top half) and their extremely luminous counterparts (bottom half) used in this study. The emission line properties in the top half refer to our Ly$\alpha$ detections, which are corrected for instrumental broadening but not for IGM or slit loss effects. Both sets of galaxies were identified by RB16 on account of their especially red \spitzer/IRAC colors and apparent brightness. All photometry is taken from RB16, while the Ly$\alpha$ properties in the lower half of the table come from other analyses. References: [1] \citet{zitrin15}; [2] \citet{oesch15}; [3] \citet{stark17}.}
\label{tab:props}
\end{sidewaystable}

\newpage

\begin{sidewaystable}[htb]
\centering
\begin{tabular}{lccccccc}
\hline
ID & $M_{\rm UV}$ & $\beta$ & log\,sSFR & Stellar Age & A$_{\rm v}$ & $Z$ & log\,$U$ \\ 
 & [mag] &  & [yr$^{-1}$] & [Myrs] & [mag] & [$Z_{\odot}$] &  \\ 
\hline
GSWY-2249353259 & -21.22$\pm$0.08 & -1.82$\pm$0.15 & -7.93$\pm$0.02 & 62.19$\pm$26.36 & 0.72$\pm$0.14 & 0.26$\pm$0.14 & -1.87$\pm$0.60\\
GSDY-2209651370 & -20.99$\pm$0.07 & -2.35$\pm$0.14 & -7.96$\pm$0.03 & 25.50$\pm$27.29 & 0.28$\pm$0.12 & 0.28$\pm$0.11 & -1.63$\pm$0.41\\
GNDY-7048017191 & -20.77$\pm$0.05 & -1.65$\pm$0.12 & -8.00$\pm$0.01 & 3.73$\pm$3.48 & 0.94$\pm$0.06 & 0.24$\pm$0.09 & -1.84$\pm$0.40\\
GNWY-7379420231 & -20.27$\pm$0.15 & -1.88$\pm$0.18 & -7.94$\pm$0.02 & 53.11$\pm$27.40 & 0.68$\pm$0.16 & 0.26$\pm$0.14 & -2.38$\pm$0.81\\
GNWZ-7455218088 & -20.57$\pm$0.15 & -1.93$\pm$0.21 & -7.94$\pm$0.02 & 43.55$\pm$26.92 & 0.64$\pm$0.16 & 0.27$\pm$0.12 & -1.98$\pm$0.63\\
\hline
EGSY8p7 & -21.99$\pm$0.06 & -1.89$\pm$0.10 & -7.94$\pm$0.02 & 46.05$\pm$27.28 & 0.72$\pm$0.10 & 0.26$\pm$0.11 & -1.61$\pm$0.37\\
EGS-zs8-1 & -22.14$\pm$0.04 & -1.98$\pm$0.09 & -7.93$\pm$0.01 & 67.05$\pm$21.22 & 0.59$\pm$0.10 & 0.30$\pm$0.13 & -1.79$\pm$0.46\\
EGS-zs8-2 & -22.05$\pm$0.04 & -2.20$\pm$0.17 & -7.98$\pm$0.02 & 12.04$\pm$20.78 & 0.47$\pm$0.14 & 0.27$\pm$0.10 & -1.59$\pm$0.35\\
COSY & -21.74$\pm$0.04 & -1.94$\pm$0.11 & -8.00$\pm$0.00 & 3.00$\pm$2.48 & 0.74$\pm$0.09 & 0.28$\pm$0.09 & -1.58$\pm$0.38\\
\hline
\end{tabular}
\caption{Galaxy properties as derived from SED-fitting of deep rest-frame UV and optical photometry with \bagpipes\ and a delayed star formation history.}
\label{tab:sed_props}
\end{sidewaystable}

\end{document}